\documentclass[aps,prc,twocolumn,showpacs,amsmath,amssymb,superscriptaddress]{revtex4}
\usepackage{graphicx}
\usepackage{dcolumn}
\usepackage{bm}
\begin{document}

\title{
Tensor interaction constraints from $\beta$ decay recoil spin asymmetry 
of trapped atoms
}
\author{J.R.A.~Pitcairn}
\affiliation{Department of Physics, University of British Columbia,
Vancouver, British Columbia, Canada V6T 1Z1}
\author{D.~Roberge}
\affiliation{Department of Physics, University of British Columbia,
Vancouver, British Columbia, Canada V6T 1Z1}
\author{A.~Gorelov}
\affiliation{Department of Physics, Simon Fraser University, Burnaby, 
British Columbia, Canada V5A 1S6}
\author{D.~Ashery} 
\affiliation{School of Physics and Astronomy, Tel Aviv University, 
69978 Tel Aviv, Israel}
\author{O.~Aviv} 
\affiliation{School of Physics and Astronomy, Tel Aviv University, 
69978 Tel Aviv, Israel}
\author{J.A.~Behr}
\affiliation{TRIUMF, 4004 Wesbrook Mall, Vancouver, British Columbia, 
Canada V6T 2A3}
\author{P.G.~Bricault}
\affiliation{TRIUMF, 4004 Wesbrook Mall, Vancouver, British Columbia, 
Canada V6T 2A3}
\author{M.~Dombsky} 
\affiliation{TRIUMF, 4004 Wesbrook Mall, Vancouver, British Columbia, 
Canada V6T 2A3}
\author{J.D.~Holt}
\affiliation{TRIUMF, 4004 Wesbrook Mall, Vancouver, British Columbia, 
Canada V6T 2A3}
\author{K.P.~Jackson}
\affiliation{TRIUMF, 4004 Wesbrook Mall, Vancouver, British Columbia, 
Canada V6T 2A3}
\author{B.~Lee}
\affiliation{TRIUMF, 4004 Wesbrook Mall, Vancouver, British Columbia, 
Canada V6T 2A3}
\author{M.R.~Pearson}
\affiliation{TRIUMF, 4004 Wesbrook Mall, Vancouver, British Columbia, 
Canada V6T 2A3}
\author{A.~Gaudin}
\affiliation{TRIUMF, 4004 Wesbrook Mall, Vancouver, British Columbia, 
Canada V6T 2A3}
\author{B.~Dej}
\affiliation{TRIUMF, 4004 Wesbrook Mall, Vancouver, British Columbia, 
Canada V6T 2A3}
\author{C. H\"ohr}
\affiliation{TRIUMF, 4004 Wesbrook Mall, Vancouver, British Columbia, 
Canada V6T 2A3}
\author{G. Gwinner}
\affiliation{
Department of Physics and Astronomy, University of Manitoba, Winnipeg,
Canada}
\author{D. Melconian}
\affiliation{
Department of Physics, Texas A\&M University, College Station, Texas, U.S.A.
}
\begin{abstract}
We have measured the angular distribution of recoiling
daughter nuclei emitted from the Gamow-Teller $\beta$ decay of 
spin-polarized $^{80}$Rb.
The asymmetry of this distribution vanishes to lowest order in
the Standard Model (SM) in pure Gamow-Teller decays, 
producing an observable very sensitive to new interactions.
We measure the non-SM contribution to the 
asymmetry to be
$A_{T}$= 0.015 $\pm$ 0.029 (stat) $\pm$ 0.019 (syst), consistent with
the SM prediction. 
We constrain higher-order SM corrections using the measured 
momentum dependence of the asymmetry, and their remaining uncertainty dominates
the systematic error.
Future progress in determining the weak magnetism term theoretically or
experimentally would reduce
the final errors.
We describe the resulting constraints on fundamental 4-Fermi 
tensor interactions.
\end{abstract}
%
\maketitle

\section{Introduction}

\subsection{Search for tensor interactions}

Effective 4-Fermi contact interactions contributing to beta decay 
can be classified by the
Lorentz transformation properties of the contributing lepton and
hadron currents~\cite{leeyang}.
The Standard Model (SM) of particle physics contains 
vector (V) and axial vector (A) interactions with sign 
`V-A'.
Extensions to the standard model can produce effective scalar and
tensor interactions. 
The observable measured here was developed shortly after the discovery
of parity violation by Treiman, who realized its sensitivity to 
tensor interactions~\cite{treiman}.

To lowest order and neglecting the Fermi function, the angular distribution
$W[\theta]$ of the daughter nuclear recoils 
with respect to the nuclear spin, integrated over all final momenta, 
is given by~\cite{treiman},

\begin{eqnarray}
W[\theta] & = &   \\
\nonumber \lefteqn{(1+\frac{1}{3}cTx_2) - x_1 (A_\beta+B_\nu) P \cos{\theta} - x_2 c T {\rm cos}^2{\theta}} \label{eq:ang} \\
P & = & \frac{\langle M \rangle }{I}~~~~~~~~~~~~~~~~~~~~~~~~~~~~~~~~~~~~ 
\label{eq:P}\\
T & = & \frac{I(I+1)-3\langle M^2 \rangle }{I(2I-1)},~~~~~~~~~~~~~~~~~~~~~~~~~~~~~~~
\label{eq:T}
\end{eqnarray}

\noindent with 
$I$ the total nuclear spin and $M$ the spin projection along the quantization
axis, so that
$P$ is the vector polarization of the nucleus, and 
$T$ is the 2nd-rank tensor alignment.
So the recoiling daughter nuclei detected in singles 
from the $\beta$ decay of polarized
nuclei have spin asymmetry, integrated over all recoil momentum,
$A_{\rm recoil}$ = --($A_{\beta}$+$B_{\nu}$)
where $A_{\beta}$ and $B_{\nu}$ 
are the $\beta$ and $\nu$ asymmetries.
The alignment coefficient $c$ depends on the nuclear spins and 
is calculated in Ref.~\cite{jtw} and Section~\ref{recoilorder}.

The coefficients $ x_1$ and $ x_2$ are calculated from integrations over the
momentum of the other outgoing particles, and their detailed dependence on
the energy release $Q$ is given in Appendix A.
In the limit of high $Q$, as is the case for our $^{80}$Rb decay,
$ x_1 \stackrel{Q>>m}{\rightarrow} 5/8$ and $x_2 \stackrel{Q>>m}{\rightarrow} 1/2$.
(Note that $x_1$ vanishes as Q $\rightarrow$ 0, as it must  
because the helicity of the $\beta^+$ vanishes as its momentum decreases to
zero.)

For Gamow-Teller decays in the SM, 
$A_{\beta}$=$-B_{\nu}$, so $A_{\rm recoil}$=0 to lowest order.
Vector and axial vector currents that couple to right-handed neutrinos 
also cancel
in the sum $A_{\beta}$+$B_{\nu}$~\cite{rhcreference}. 
That makes $A_{\rm recoil}$ sensitive only to fundamental
lepton-quark tensor interactions: it is neither 
sensitive to effective scalars nor to right-handed currents. 
The exclusive sensitivity to tensor interactions is only true for one other 
observable, the $\beta$-$\nu$ correlation
in pure Gamow-Teller decay~\cite{ornl,gluck}. 

The near-zero value of $A_{\rm recoil}$
in the allowed approximation for 
pure Gamow-Teller decays also
makes it a very attractive experimental observable; for example, 
the polarization P
does not have to be known to high precision to extract the new physics. 
We will see below that recoil-order corrections produce effects $\sim$ 0.01 
in the absence of new physics.
Note from Appendix A that the recoil spin asymmetry remains proportional to 
$A_{\beta}$+$B_{\nu}$ even before the integration over recoil momenta.

\subsubsection{Explicit sensitivity to 4-Fermi tensor coupling coefficients}

The tensor-axial vector Fierz interference term is linear in a particular
combination of the 4-Fermi tensor coupling coefficients 
$C_T$ and $C_T'$~\cite{jtw}:

\begin{equation}
b_T\frac{m}{E_{\beta}} = 
\frac{(C_T+C_T')}{C_A}\frac{m}{E_{\beta}}.
\label{eq:bt}
\end{equation}

\noindent This term appears in the full 
expression for
the decay distribution before integration over the $\beta$ momentum~\cite{jtw}.
It is well-constrained by other experiments in 
lower-energy $\beta$ decays~\cite{carnoy}, so we will cite results below
both with it unconstrained and with it constrained to be small.
Its inclusion complicates the expressions, so we include it in our
full simulations by numerical
integration; we also show the full expression (without Fermi function)
in Appendix A.
Note that Eq.~1
has consciously 
assumed the Fierz term to be zero~\cite{treiman}
to simplify the expressions.

In absence of this Fierz term, 
the contribution to $A_{\rm recoil}$ from non-SM interactions 
becomes just the product of 
the 4-Fermi tensor constants 
$C_T$ and $C_T'$~\cite{treiman}:

\begin{equation}
A_{\rm recoil} \stackrel{b_T=0}{\rightarrow} A_{T} =  \pm 2 \lambda_{I,I'} C_T C_T'/C_A^2 
\label{eq:Arecoil}
\end{equation}

\noindent where the $\pm$ sign is for $\beta^{\pm}$ decay.
The coefficient $\lambda_{I,I'}$ is given in Ref.~\cite{jtw} 
(see Section~\ref{recoilorder}). 

The combination $C_T+C_T'$ describes a tensor interaction 
that couples to standard model left-handed neutrinos, while the coupling
$C_T-C_T'$ describes an interaction  coupling to neutrinos with 
non-SM helicity~\cite{severijns}.
So the sensitivity of $A_{\rm recoil}$ to $C_T C_T'$ 
produces sensitivity to interactions with both SM and
non-SM chirality.

\subsection{Tensor interactions in other experiments and theory}

Until recently, individual 
nuclear $\beta$-decay correlation experiments and global fits 
have been consistent with 
the standard model without 
a tensor interaction~\cite{herczeg2,voytas,skalsey}.
A recent global fit  
of nuclear and neutron $\beta$ decay data 
including scalar and tensor terms 
coupling only to standard-model left-handed $\nu$'s (i.e. assuming
$C_T$=$C_T'$) gives
$C_T/C_A$ = 0.0086$\pm$0.0031, while
excluding the latest neutron lifetime 
measurement~\cite{serebrov} from that fit brings the result into agreement
with the Standard Model at one standard deviation~\cite{severijns}.
Such difficulties in combining many experiments with different systematic
errors in global fits can be avoided by dedicated experiments
sensitive only to tensor interactions, like the present experiment.
So the present measurement becomes useful if it achieves $\sim$ 0.01 accuracy 
in the recoil asymmetry.  

A tensor interaction
coupling to right-handed neutrinos would produce a contribution to the
mass of the standard model left-handed neutrinos.
An order-of-magnitude calculation suggests that if 
$|C_T-C_T'|$ were $\sim$ 0.02, such an interaction would account for neutrino
masses $\sim$ 3 eV~\cite{ito}, the present experimental upper limit.
This provides a motivation for direct correlation measurements in
order to constrain this possible contribution to neutrino masses.

The PIBETA collaboration reported 
a statistically significant 
deviation from the standard model~\cite{Frlez1}
in $\pi$$\rightarrow$$\nu$$e$$\gamma$ decay that could be explained by
a finite tensor interaction.
The same group has made further dedicated experiments and now eliminated the
possibility of a tensor interaction~\cite{bychkov}.
The most restrictive limits from
$\pi$$\rightarrow$$\nu$$e$$\gamma$ decay are on interactions with same
chirality as the Standard Model, i.e. constraining a Fierz interference term 
that is linear in the small tensor term, 
although the opposite chirality is also
considered in Refs.~\cite{herczeg2,poblaguev}.
Sensitivity in the present experiment $\sim$ 0.001 level
would be needed to reach the sensitivity probed
by the $\pi$ decay experiments. 

A renormalizable tensor interaction 
can be generated by the exchange of 
spin-0 leptoquarks~\cite{herczeg1,herczeg2}.
In an example of an explicit model, 
a recent analysis of the possible one-loop corrections in 
SUSY models shows they can produce tensor and scalar interactions as large as 
0.001 in the Fierz interference term $b_{T}$ from
left-right sfermion mixing in the first generation, 
physics that is otherwise difficult to constrain~\cite{profumo}.

\subsection{$^{80}$Rb and recoil-order corrections}
\label{recoilorder}

\begin{figure}
\includegraphics[angle=0,width=0.8\linewidth]{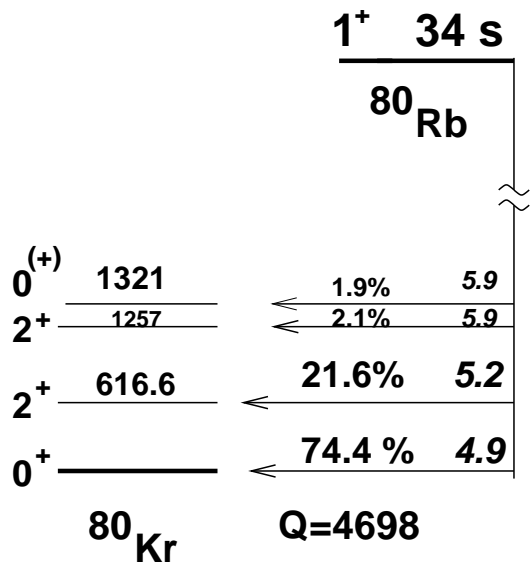}
\caption{ 
The $\beta^+$-decay scheme for $^{80}$Rb, showing literature values of the
spin, parity, branching
ratio, log$_{10}$(ft), and excitation energy of the final levels in $^{80}$Kr,
along with the Q-value (maximum kinetic energy) for $\beta^+$ decay and
the parent half-life. Energies in keV.
}
\label{Rb:nuclearlevels}
\end{figure}

The decay of I$^\pi$=1$^+$ $^{80}$Rb 
is primarily to two states, 74\% to the ground 0$^+$ state and 22\% to
the first excited 2$^+$ state (see Figure~\ref{Rb:nuclearlevels}).
The coefficient $\lambda_{I,I'}$ in Eq.~\ref{eq:Arecoil} 
is 1 for the 1$^+$$ \rightarrow$0$^+$
transitions and -1/2 for
the 1$^+$$\rightarrow$2$^+$ transitions. So when averaged over the transitions,
the average $\lambda_{I,I'}$ for $^{80}$Rb is 
0.64.
The coefficient $c$ in Eq.~1 is
1 for the 1$^+$$\rightarrow$0$^+$ transitions, and 
+0.1 for 1$^+$$\rightarrow$2$^+$.

The 1$^+$$\rightarrow$0$^+$ transition has higher-order corrections in 
the SM proportional to the 
weak magnetism form factor $b_M$ and the 
induced tensor form factor $d$~\cite{holstein}.
The contributions to the angular correlation coefficients 
from these recoil-order terms are scaled by $E_{\beta}/M_{\rm nucleon}$, 
and they produce a small nonzero 
recoil asymmetry within the Standard Model.

These recoil-order terms are given by sums over individual nucleon matrix
elements

\begin{eqnarray}
b_M/A & = & g_M M_{GT} + \langle f || \sum_{k} \tau_k^+ \vec{l}_k || i \rangle  
\\
d/A & = & g_A \langle f || \sum_k \tau_k^+ i \vec{\sigma}_k \times \vec{l}_k || i \rangle
\label{eq:bd}
\end{eqnarray}
using the notation of Ref.~\cite{chung}, which has detailed consideration of the recoil-order matrix elements needed here.

The first  
term in the expression for weak magnetism, $b_M/A$, is 
from the anomalous isovector magnetic moment of the nucleons, 
and is given by the dimensionless number $g_M$=4.7 multiplying the Gamow-Teller
matrix element $M_{GT}$. 
The second term in $b_M/A$ 
requires a detailed nuclear structure calculation with a 
large fractional uncertainty in this mass region, as deformation effects
make shell model calculations difficult.
Weak magnetism terms for a variety of Gamow-Teller decays have been calculated
and measured, and the values generally do not depart greatly from the value
of $b_M/(AM_{GT})$=4.7~\cite{chung,holstein}.

The induced tensor term $d$ is more poorly characterized. We will see below
that it contributes values to the experimental recoil asymmetry 
that are roughly constant
with recoil momentum, so we will be able to fit for $d$ simultaneously with the
non-SM tensor interaction. 

\subsection{Transitions to excited states}
The 22\% 1$^+$ to 2$^+$ transition in principle 
adds some complication to the recoil-order matrix elements.
The first-order form factors, $b_M$ and $d$, can be different from those
for the transition to the ground state. In addition, 
a total of five additional form factors appear at 2nd-order in recoil terms
$E_{\beta}/M_{\rm nucleon}$.
In this version of the experiment,
we will let $b_M$ and $d$ float phenomenologically below, so that 
we implicitly include $b_M$, $d$ for the excited state in our analysis.
In other words, $b_M$ and $d$ for both the ground
and excited states are treated as producing the same functional dependence of
the recoil asymmetry on recoil momentum. This is an excellent
approximation as the effects on the recoil asymmetry have very similar momentum
dependence compared to the ground state, particularly once they are
averaged over the momentum spread induced by the final $\gamma$-ray emission.
The values of $b_M$ and $d$ that are extracted are then weighted averages of
the transitions.  
We ignore the 2nd-order terms in E$_{\beta}$/$M_{\rm nucleon}$, which is a 
good approximation at the level of accuracy reached in this version of the 
experiment.

\section{Experimental Techniques}

We describe below the trap apparatus which we use to polarize the
$^{80}$Rb nuclei and measure the angular distribution and 
momentum of the $^{80}$Kr daughter.

The collection and trapping of the $^{80}$Rb 
used a two-trap apparatus~\cite{swanson} very similar to that
in previous $\beta$-decay work~\cite{gorelov,melconian}.
The $^{80}$Rb 30 keV ion beam from the ISAC facility at TRIUMF 
was $\approx$ 2 $\times$ 10$^9$/sec from a zirconium carbide target.
The ion beam was stopped in a neutralizing foil made of zirconium.  
The resulting atoms were collected in the first magneto-optical trap 
(MOT)~\cite{raab}. The trapped atoms were transferred to the second
MOT to minimize backgrounds~\cite{swanson} 
and provide an environment for polarization. 
The number of atoms continuously trapped in the detection trap 
was $\approx$ 2 $\times$ 10$^6$. 

\subsection{Detection geometry} 
 
\begin{figure}
\includegraphics[angle=0,width=0.9\linewidth]{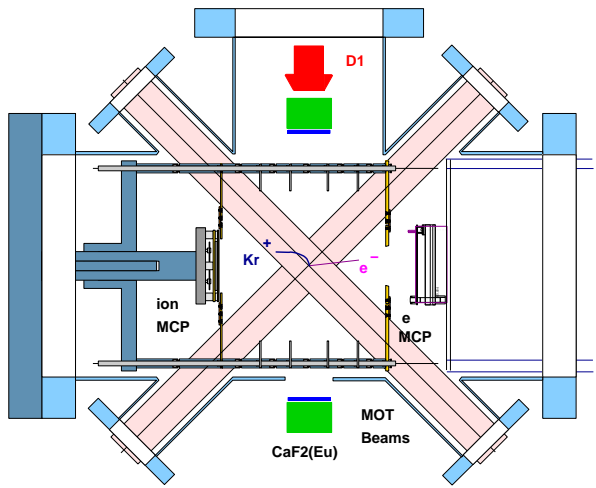}
\caption{ (Color online)
TRINAT detection apparatus. The MCP for ion detection
has position-sensitive resitive anode readout. 
Given the known uniform electric field, measurement of 
the recoil ion TOF and impact position
determines the initial momentum.
The apparatus is similar to that of~\cite{gorelov}, with
an MCP for electron
detection added for the $^{80}$Rb decay asymmetry.
The $^{80}$Rb is spin-polarized by optical pumping with  
light resonant to the D1 transition from a 50 mW diode laser. 
The polarization is monitored with
$\beta$ $\Delta E-E$ phoswiches using plastic/CaF$_2$(Eu) fast/slow
scintillators (which are at --30 and 150 degrees out of the plane shown here).
}
\label{Rb:apparatus}
\end{figure}

The detection geometry is shown in Fig.~\ref{Rb:apparatus}.
The time-of-flight (TOF) 
of the daughter nuclei from nuclear $\beta$ decay 
in singles (i.e. not in coincidence with the $\beta$) 
can be measured by using the atomic
shakeoff electrons as a trigger, as developed by LBL researchers
~\cite{vetter}. 
A uniform electric field of average value 800 V/cm 
collects ions produced in $\beta$ decay to  
a 25 mm diameter microchannel plate (MCP) Z-stack for time readout, backed by 
a position-sensitive resistive anode.
The electric field also collects the atomic electrons to 
a second MCP detector on the opposite side.
A positive ion
produced in $\beta^+$ decay will `shake off' at least two atomic electrons, 
and electrons
up to $\sim$100 eV energy are completely collected by the field into the 
detector.
So this technique increases
the efficiency for recoil detection by a factor of about 30 compared to
$\beta^+$ detection (which had solid angle 1\% in our geometry~\cite{gorelov}).
This produced approximately 100 Hz of daughter recoils
in coincidence with the electron detector, for the average number of 
atoms trapped of about 2$\times 10^6$.

The electron detection geometry and electric fields are optimized to 
ensure efficient detection independent of the electron's 
initial kinetic energy and angle. 
We measured $\approx$ 35\% detection efficiency for 
low-energy electrons
from the laser photoionization of $^{80}$Rb (Fig.~\ref{Rb:D1}). 
By photoionizing stable $^{85}$Rb, we have 
reproduced data in the literature showing that the electron efficiency 
of MCPs is optimized at 500 eV impact energy, and changes by less than 10\% 
between 500 and 1000 eV ~\cite{mcpelectron}. The shakeoff electrons are
expected to have kinetic energies similar to their atomic binding energies,
a few 10's of eV~\cite{shakeoff},
so we arrange the fields so that
the electrons impact the MCP with approximately 500 eV 
more than their original kinetic energy.
To make the electric field
more uniform in the region traversed by the electrons than 
in the geometry of Ref.~\cite{gorelov}, 
the grid in front of the electron
detector is biased 
to a potential close to that of the final electrode in the field
assembly.

\subsection{Polarization techniques}

\begin{figure}
\includegraphics[angle=0,width=0.6\linewidth]{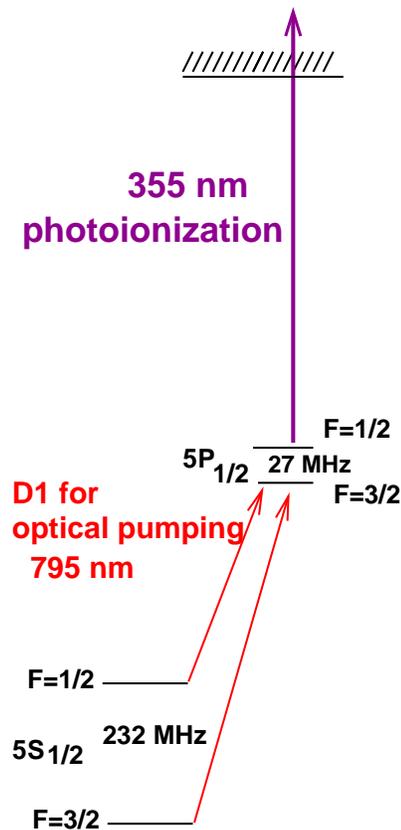}
\caption{ (Color online)
Energy diagram of the
atomic levels for optical pumping of the Rb atom, along with the 
photoionization scheme for position determination.
The atomic hyperfine structure is not to scale.
The frequency splittings were taken from ~\cite{orsay}.
}
\label{Rb:D1}
\end{figure}

The $^{80}$Rb atoms are polarized by switching off the MOT light and
optically pumping with light at the D1 transition (see Fig.~\ref{Rb:D1})
for 30 $\mu$s. Then the MOT light is switched on again for
30 $\mu$s to keep the atom cloud from expanding.  
The MOT 3 G/cm (horizontal) quadrupole field stays on at all times.
In a typical MOT with beams carefully balanced in power, 
the atom cloud would be centered at zero
magnetic field, but such a cloud would sample nonzero fields and the 
polarization would be disturbed by Larmor precession.
In order to add a constant magnetic field along the
optical pumping axis, we attentuate two of the beams in the
MOT horizontal plane, perpendicular to the quadrupole field anti-Helmholtz
coil axis. The cloud equilibrium position is then at finite B
field,
i.e. at a location where Zeeman shifts produce equal absorption from the
unbalanced beams.   
Then we in addition 
apply at all times a uniform field constant of 2.5 G with Helmholtz
coils. The result is a cloud equilibrium position at the center of the 
apparatus, with on average a 2.5 G field along the optical pumping axis. 
The cloud spatial FWHM of 3 mm samples $\pm$0.5 Gauss of 
changing field.

Atoms were transferred from the 1st trap every 1.5 seconds.
After each transfer, the polarization state was flipped by changing the
handedness of the optical pumping light with a liquid crystal variable
retarder. 
 
The 30 $\mu$s trap on/off polarization off/on duty cycle was chosen to
minimize the motion of the cloud during the optical pumping time, which 
was a dominant systematic in $\nu$ asymmetry measurements 
in $^{37}$K~\cite{melconian}.
The 2.5 G bias B field means that there were slightly 
different Zeeman shifts for
the two different polarizations.
To minimize differences in the atom cloud position,
the optical pumping laser frequency was shifted (by less than a linewidth)
between the two polarizations.  

The atom cloud position was monitored by photoionizing a small fraction
of the atoms with a pulsed laser (see Fig.~\ref{Rb:D1}) and recording
MCP position and TOF. 
The average trap location was found to shift by 0.030$\pm$0.003 mm 
with spin flip,
which would produce a false asymmetry A$_{\rm recoil}$ of 0.0012$\pm$0.0001.
The correction was made to the data by
using the measured trap location when determining the experimental angle of
emission below (Section~\ref{analysis}).

\subsection{TOF spectrum}
\label{subsection:electron}

\begin{figure}
\includegraphics[angle=90,width=0.8\linewidth]{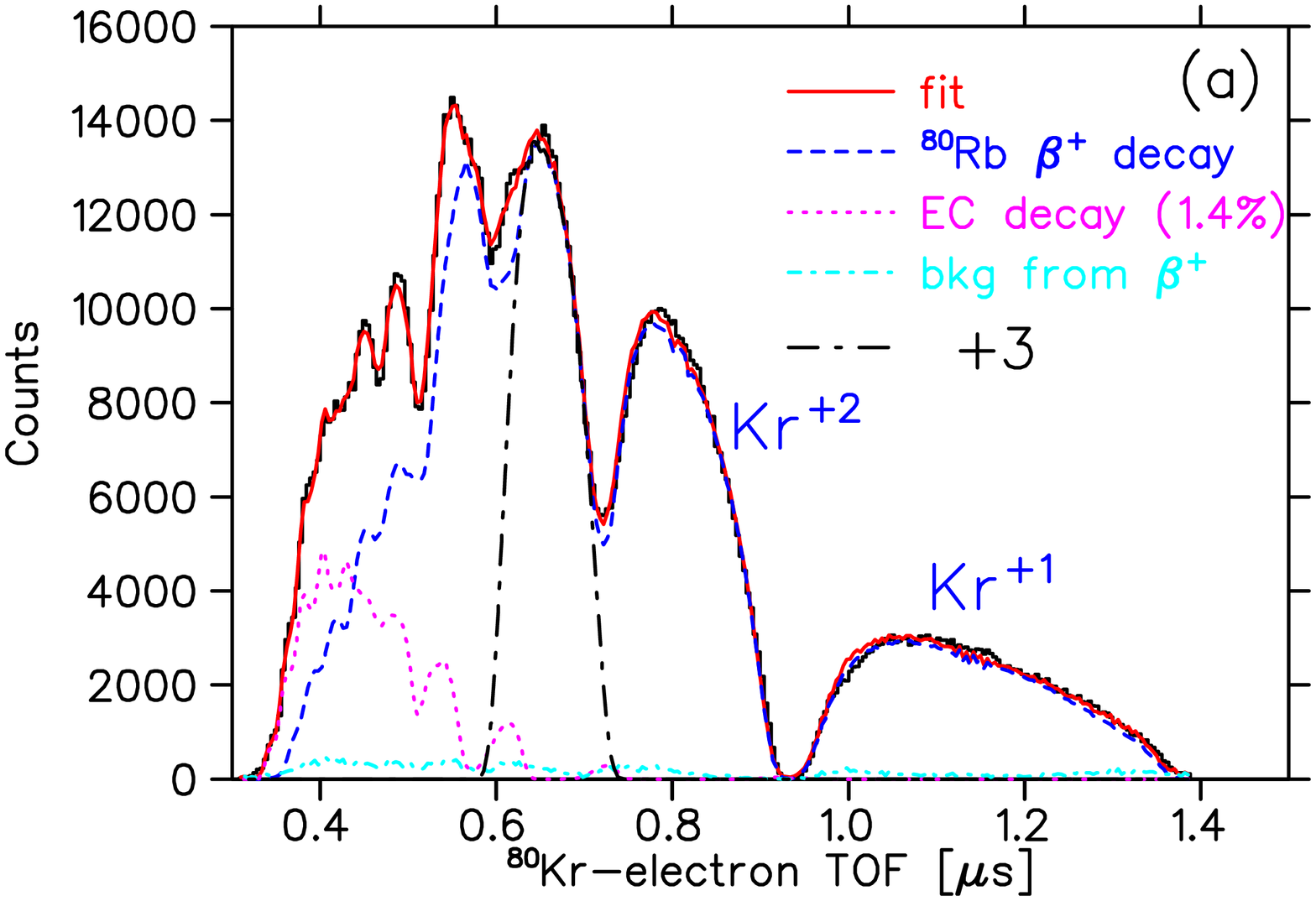}\\
\includegraphics[angle=90,width=0.8\linewidth]{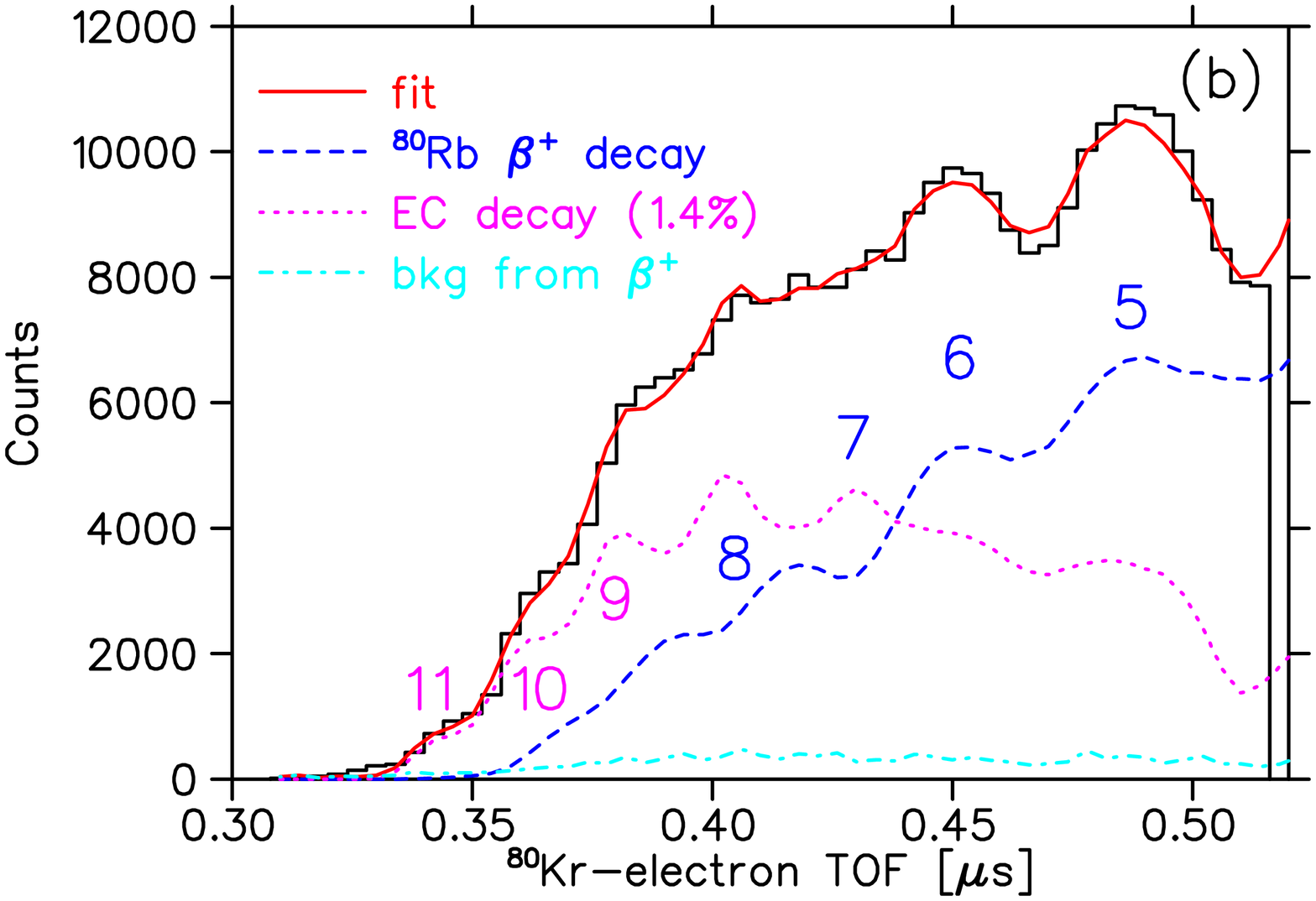}
\caption{ (Color online)
(a)
Time-of-flight (TOF) spectra for recoil coincidences with 
shakeoff atomic electrons, showing decomposition into $\beta^+$ decay
and backgrounds. The small electron capture (EC) 
branch produces large corrections
for the higher charge states, and is modelled here assuming charge state
distributions from x-ray photoionization~\cite{carlsonhunt}. 
A background of order 1\% from $\beta^+$'s striking the
electron detector is determined by biasing the detector to exclude low-energy
electrons (see text).
The simulation for charge state +3 is highlighted; 
the TOF for +3 was cut above 0.63 $\mu$s to help exclude EC events.
(b)
Expanded TOF scale of the top figure, showing 
the charge states 9, 10, and 11 that are dominated by 
EC and used for polarization determination 
(Section~\ref{sec:pol}).
}
\label{Rb:tof}
\end{figure}

We show a typical TOF spectrum in Fig.~\ref{Rb:tof}, 
deconstructed into its components.
The different charge states are separated in TOF by a uniform
0.8 kV/cm electric field. Charge states 1, 2 and most of charge state 3 
are relatively clean of
background. Ions from the 1.4\% electron capture branch have a 
large spin asymmetry and
contaminate the higher charge states, and because their asymmetry is large
they become a useful probe of the polarization (see below). 
Note that only about 15\% of the
$\beta^+$ decays produce positive ions, as opposed to most of
the electron capture decays. 

We measured the background from $\beta^+$'s 
striking the electron detector by lowering the bias voltage of the
detector to exclude the atomic electrons, while keeping the grid in
front of the detector at the same voltage to keep the electric field
for ion collection the same.
That background has
the expected large spin asymmetry from $\beta$-recoil coincidences, which in 
this geometry is determined by the $\nu$ asymmetry~\cite{melconian}. 
The recoil asymmetries shown below have been corrected for this background, 
which produces 1.2\% of the charge state 1 recoils, 
0.59\% of the charge state 2 recoils, and 0.51\% of the charge state 3 recoils.
The average asymmetry correction can be seen below in Section~\ref{analysis},
Fig.~\ref{Rb:withwithoutbeta} and will be discussed there.
  
There is in addition a 1\% background between 1.2 and 1.4 $\mu$s that shows
a definite localization on the lower
part of the ion MCP. This background also appears in natural backgrounds
and $\gamma$-ray source measurements, and may be due to an electronic
artifact.
Although its origin is not fully understood, we have measured its spin 
asymmetry to be negligibly small and consistent with zero, 
so via this technique we can correct for its presence to sufficient accuracy.

\subsection{Vector polarization determination}
\label{sec:pol}

Here we describe the determination of 
the nuclear vector polarization achieved, 
0.55$\pm$ 0.04. 
The precision is more than adequate because the present observable vanishes,
though the final error on the new physics parameters $A_T$ and $b_T$ is
compromised because it 
scales inversely with the absolute polarization achieved.

The polarization was optimized by measuring the time dependence of the atomic 
excited state population during the optical pumping, monitored by non-resonant
photoionization with a small pulsed laser (Fig.~\ref{Rb:D1}). 
The excited state population 
decreases as the polarization increases: if the atoms become fully polarized,
then the atom can absorb no more light in this transition and the 
excited state population would vanish.
Equilibrium polarization is reached
during the last 20 $\mu$s of the optical pumping,
during which $A_{\rm recoil}$ is measured.
Unlike in our previous work with the MOT magnetic quadrupole field turned 
off~\cite{melconian},
we found that the atomic measurement of the polarization was difficult
to quantify with the MOT quadrupole field left on.
So the polarization was 
measured by nuclear observables.

The $\beta^+$ asymmetry was measured using plastic/CaF$_2$(Eu) 
phoswich detectors, in coincidence with shakeoff electrons to minimize
sensitivity to decays from untrapped atoms.
The phoswiches are located at --30 and 150 degrees with
respect to the polarization direction, out of the plane of 
Fig.~\ref{Rb:apparatus}.

\begin{figure}
\includegraphics[angle=90,width=0.8\linewidth]{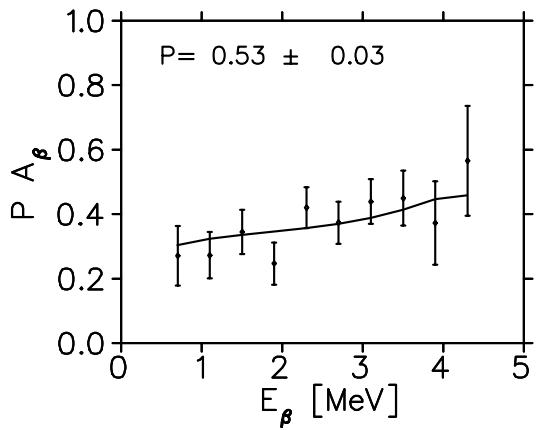}
\caption{
Nuclear polarization determined from the $\beta$ asymmetry in coincidence
with Kr recoils charge states 4-9
in the CaF$_2$(Eu) detectors.
}
\label{Rb:betaasymmetry}
\end{figure}

The $\beta$ asymmetry for the 1$^+$ to 0$^+$ transition is +1, 
while for the 1$^+$ to 2$^+$ it is -1/2.
So the asymmetry grows at the higher $\beta$ energies, as the contribution
of the 1$^+$ to 2$^+$ becomes proportionately smaller. 
Figure~\ref{Rb:betaasymmetry} shows
the fit asymmetry as a function of $\beta$ momentum.
For betas in coincidence with
recoils in charge states +4 to +9, the full solid angle of the recoils is
detected, so this physical observable is the same as the singles
$\beta$ spin asymmetry. It has the advantage of being completely clean
of background from decaying atoms that are not in the trap.
The result is P=0.53 $\pm$ 0.03.

\begin{figure}
\includegraphics[angle=90,width=0.8\linewidth]{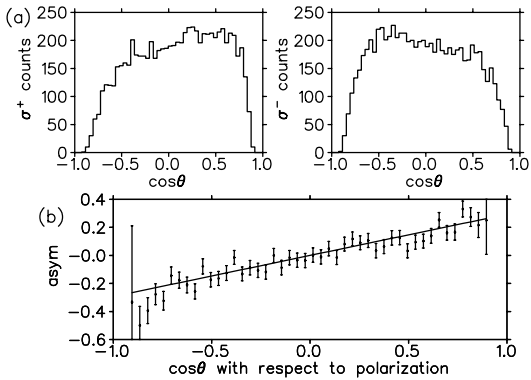}
\caption{
Nuclear polarization determined from recoils from electron capture (EC) decay
(see text), which has a large spin asymmetry.
(a) Counts as a function of cos~$\theta$ for opposite signs of polarization.
(b) The resulting asymmetry of events from (a); the line is a fit
with $P A_{EC}$=0.29$\pm$0.02 (see text). 
}
\label{Rb:ECasymmetry}
\end{figure}

An entirely different nuclear observable serves as an additional measurement
of the nuclear polarization. Recoils produced in electron capture (EC)
decay dominate the higher charge states 9, 10, and 11 
(see Figure~\ref{Rb:tof}(b).)
A further cut on recoil momentum removes a 25\% contribution from
$\beta^+$-produced recoils, leaving a
clean sample of the highest-momentum EC-produced recoils.
Figure~\ref{Rb:ECasymmetry}(b) shows the angle dependence (as constructed from
the MCP position information and time-of-flight)
of the resulting recoils from EC. If the polarization $P$
were unity, the EC recoils would have asymmetry unity for the 
1$^+$ to 0$^+$ transition, and -0.5 for
the 1$^+$ to 2$^+$ transition. A simple linear fit to the asymmetry in 
Figure~\ref{Rb:ECasymmetry}(b) 
(there is no cos$^2$$\theta$ term~\cite{treiman}),
extracts $PA_{EC}$=0.29$\pm$0.02, which implies 
nuclear polarization $P$=0.57$\pm$0.04. 

In summary, the nuclear vector polarization achieved from these consistent
observables is 
0.55 $\pm$ 0.04. 
The uncertainty 
produces an error on the extracted $A_{\rm recoil}$ of 7\% of its value, which
as we will see below is negligibly small in this version of the experiment.

\subsection{ Tensor alignment}
We have no direct observables that are very sensitive to the tensor-order
polarization alignment $T$ (defined in Eq.~\ref{eq:T}), 
but we can make adequate indirect 
constraints from the
measured vector polarization $P$=0.55$\pm$0.04.
For nuclear spin $I$=1 and $P$$<$1, the population that is not in
spin projection $m_I$=1 must obviously either 
be in $m_I$= 0 or -1. If it were all in $m_I$=-1, then the
value of $T$=--1, unchanged from its value for perfect $P$=1.
For imperfect optical pumping spoiled by Larmor precession in the 
quadrupole field, it is much more likely for most of the population to
be in $m_I$=0; doing this produces our best estimate of $T$=0.35.
In the fits, these two extremes do not 
significantly perturb the extracted asymmetry coefficients $A_1$, so
we do not mention them further, and simply take $T$=0.35 in the remaining
analysis.

\section{Analysis of recoil asymmetry as a function of recoil momentum to extract recoil-order terms}
\label{analysis}

From the ion MCP hit position, ion TOF, trap cloud location from 
photoionization, known uniform electric field, and known charge states 
1-3 from range of TOF,
we can construct the momentum of the recoil and its 
emission angle with respect to the polarization direction.
By fitting the angular distribution of the recoils for different momentum 
bins, we can extract the recoil asymmetry as a function
of recoil momentum. 

The dependence on recoil momentum 
allows us to extract information about the 
recoil terms while simultaneously fitting for the tensor interaction. 
These different terms produce a different functional dependence of the 
recoil asymmetry on
momentum.

\begin{figure}
\includegraphics[angle=90,width=0.8\linewidth]{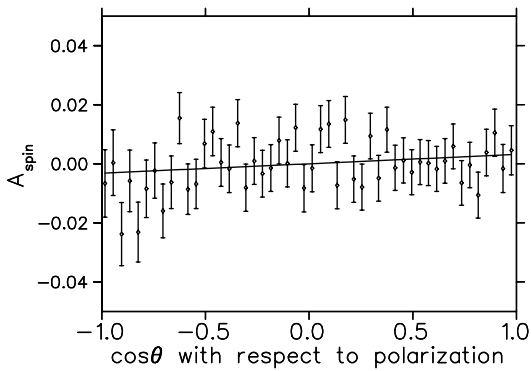}
\caption{
An example of the recoil asymmetry fit as a function of angle, with angle
reconstructed from the impact location and the ion TOF. Here the 
asymmetry is for the charge state 2 data, summed over all
recoil momentum.
}
\label{Rb:cosasym}
\end{figure}

If we were to integrate over all momenta,
the measured asymmetry $A_{\rm spin}$ as the spin polarization 
P is flipped can then be found from Eq.~1:\\

\begin{eqnarray}
A_{\rm spin} & = &
\frac{W[\theta, P] - W[\theta, -P]}{W[\theta, P] + W[\theta, -P]}\\
\nonumber &  = &
\frac{x_1 P A_{\rm recoil} {\rm cos}{\theta}}{ 1 + cTx_2+ c T x_2 {\rm cos}^2{\theta}}
\label{eq:aspin}
\end{eqnarray}

\noindent still ignoring recoil-order terms.
An example of such a fit that could extract the quantity 
$P A_{\rm recoil}$ is
shown in Fig.~\ref{Rb:cosasym}.

We wish to generalize this to a fit of the recoil asymmetry as a function
of recoil momentum. In the absence of the Fermi function and of recoil-order
corrections, 
we could define
the recoil asymmetry as a function of recoil momentum $A_{\rm spin} [P_r]$
in terms of the kinematic functions of Appendix A:

\begin{eqnarray}
A_{\rm spin}[P_r] =
\frac{W[\theta, P, P_r] - W[\theta, -P, P_r]}{W[\theta, P, P_r] + W[\theta, -P, P_r]} =\\
\nonumber \frac{(f_4 A_{T} - f_7b_T) P {\rm cos}{\theta}}
{ f_1 -b_T f_6 -(a_{\beta\nu} + \frac{cT}{3})f_2 
+cT(f_3+f_5{\rm cos}^2(\theta))}
\label{eq:aspinf}
\end{eqnarray}

\noindent where all the $f_i$ functions depend on 
$P_r$.
However, we must properly include the Fermi function (which does not matter
quantitatively) and the recoil order terms (which do matter).
So instead we write the simple expression actually used for the fits
as a function of angle, for each bin of recoil momentum:

\begin{eqnarray}
A_{\rm spin}[P_r] & = &
\frac{P A_1[P_r] {\rm cos}{\theta}}{ 1 + c T F_2[P_r] {\rm cos}^2{\theta}}
\label{eq:a1}
\end{eqnarray}

\noindent where $F_2[P_r]$ now contains all the numerical integrations needed for
the cos$^2$($\theta$) terms. 
These fits let us extract the experimental 
coefficient of cos($\theta$) of the recoil
asymmetry,  $A_1[P_r]$. We then will fit $A_1[P_r]$ below, using the 
different dependence on recoil momentum of the new tensor physics terms and
the recoil order terms, in order to extract the
tensor physics terms $A_T$ and $b_T$. 

We show one experimental 
example of the spin-flip asymmetry $A_{\rm spin}[\theta]$ 
as a function of 
cos($\theta$) in Fig.~\ref{Rb:cosasym}.
We make similar fits to $A_{\rm spin}$ as a function of binned recoil
momentum from 0.5 to 5.0 MeV/c to extract $A_1[P_r]$.

\subsection{Experimental results for $A_1[P_r]$}

\begin{figure}
\includegraphics[angle=0,width=0.8\linewidth]{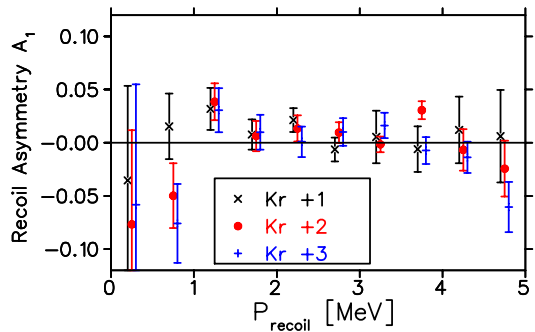}
\caption{ (Color online)
The dependence on recoil momentum of the experimental recoil asymmetries $A_1$.
Charge states 1,2, and 3 are shown to be in statistical agreement.
(The points for charge states 1 and 3 are offset horizontally for clarity.)
}
\label{Rb:asympch123}
\end{figure}

We show the results for $A_1[P_r]$ broken down for the different charge states 
in Figure~\ref{Rb:asympch123}.
The first three charge states can be seen to be statistically consistent.
We therefore simply take the weighted average over the charge states to 
consider the physics below. 

\subsection{Correction for $\beta^+$'s striking MCP}
\begin{figure}
\includegraphics[angle=0,width=0.8\linewidth]{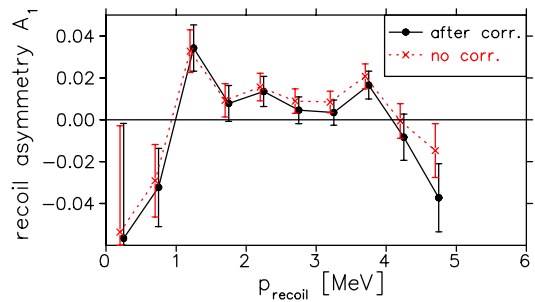}
\caption{ (Color online)
$A_1$ as a function of recoil momentum, determined with no correction for
$\beta$'s and 
after the correction for $\beta^+$'s described in Section~\ref{subsection:electron}. The correction is small. 
Lines are drawn only to direct the eye.
}
\label{Rb:withwithoutbeta}
\end{figure}

Since there is generally concern in precision $\beta$ 
asymmetry experiments with the size of any corrections, we show in 
Figure~\ref{Rb:withwithoutbeta} the results for 
$A_1$ with and without the 
correction from $\beta^+$'s striking the electron MCP, as described in
Section~\ref{subsection:electron}. The final answer for the average
$A_{\rm recoil}$, shown below, becomes more negative by about 0.01 when
the $\beta^+$ correction is made. We know this 
correction to better than 5\% of its value.

\section{Results of fits to the experimental $A_1$}
\label{results}

We now show results for fits to the extracted average $A_1[P_r]$. 
We try to take advantage of the  
different momentum dependence produced by the
recoil-order corrections $b_M$ and $d$, 
and the non-SM tensor physics parameters
$A_T$ and $b_T$. We consider different constraints on the parameters, 
partly in hopes that in the future values of the recoil-order 
corrections may be available. 

\subsection{All parameters floating}

\begin{figure}[htb]
\includegraphics[angle=90,width=0.8\linewidth]{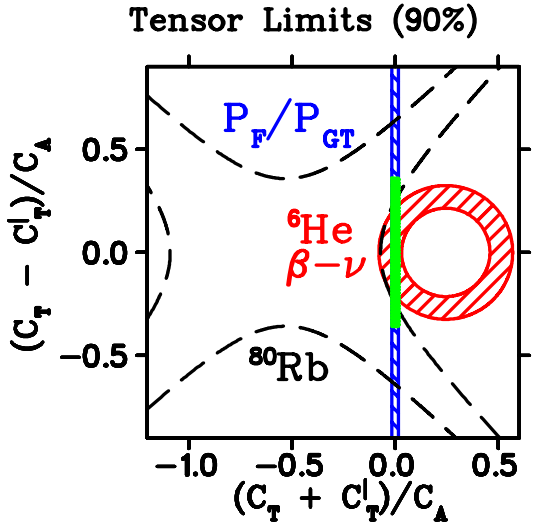}
\caption{ (Color)
Exclusion plot at 90\% confidence 
showing complementarity of the present constraints to other 
measurements. Allowed regions are 
inside: 
$^6$He $\beta$-$\nu$~\cite{ornl,gluck},
red concentric circles (hashed in between);
$\beta^+$ polarization in $^{14}$O, $^{10}$C~\cite{carnoy}
(assuming no scalar interaction),
blue solid vertical lines (hashed in between); 
Present work, letting $b_M$, $d$, and both $A_T$ and $b_T$ float,
black dashed hyperbolae.
For $b_T$ set to 0 (consistent with~\cite{carnoy}),
weak magnetism $b_M/(AM_{GT})$= 4.7 $\pm$ 4.7, and $d$ left floating, then
the green rectangular area 
indicates the present limit $|(C_T-C_T')/C_A|<0.36$. 
}
\label{Rb:exclusionLR}
\end{figure}

First we let all parameters float. We consider various 
points in the $C_T+C_T'$ vs. $C_T-C_T'$ plane, let $b_M$ and $d$ float for 
each point, and compute the $\chi^2$. We show the 90\% CL limits in a 
contour plot in Fig.~\ref{Rb:exclusionLR}, including constraints
from other experiments. 
The best fit is for
$b_M/(A M_{GT})$=-7.2$\pm$5.0, a negative value with rather large absolute
value, and a small
value of $d/A$=0$\pm$17. 
The square root 
of the reduced $\chi^2/N$ 
is 1.37 for the best fit, so we expanded the error bars on the fit parameters 
appropriately. 

\begin{figure}[h]
\includegraphics[angle=0,width=0.8\linewidth]{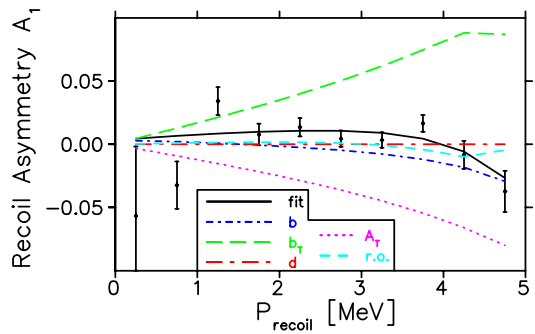}
\caption{ (Color online)
The dependence on recoil momentum of the recoil asymmetries for the best
fit with $b_M$, $d$ and both chiralities of tensor floating. 
$\sqrt{\chi^2/N}$=1.37 for this fit.
The cyan medium-dash line `r.o.' is a small exact recoil-order correction
given when recoil energy is included in the energy conservation equation.
}
\label{minplot}
\end{figure}

As a tool to discuss this result, we show the individual 
contributions from
$b_M$, $d$, $A_T$, and $b_T$ in Fig.~\ref{minplot}. The momentum dependence
of $A_1$ from the non-SM tensor terms $A_T$ and $b_T$ is similar, so
they are both made large with opposite signs to fit the data. The asymmetries
produced by these non-SM tensor terms in this fit 
are therefore much larger in absolute
magnitude than the experimental asymmetries. 

So a completely unconstrained fit does not produce 
competitive limits in a model with all chiralities of tensors possible and
no constraints on the recoil order terms,
because of the large number of similar degrees
of freedom. It is also clear that the statistical precision of the data would
be much better than the size of these error bars, if other constraints on the
physics were applied.

\subsection{Constraining $C_T+C_T'$=0}

We next use the 
relative positron polarimetry experiments comparing pure Fermi 
$^{14}$O and the pure Gamow-Teller $^{10}$C branch 
~\cite{carnoy} to imply that the
Fierz interference term $b_T$ $\propto$ $C_T+C_T'$ is very small.
We continue to let $b_M$, $d$, and $A_T$ float.

The resulting constraints on 
$C_T-C_T'$ can simply be seen by looking at the intersection of our
limit hyperbolae with the
the $C_T+C_T'$=0 axis of Fig.~\ref{Rb:exclusionLR}.
A more detailed look at the $\chi^2$ minimum implies 
$|(C_T-C_T')/C_A|$=0.42$^{+0.15}_{-0.32}$ at 90\% confidence, and the
resulting 
recoil order terms are 
$b_M/(AM_{GT})$=-18$\pm$11 and $d/A$=24$\pm$41. Note here that $b_M$ lies
far outside the nucleon value.

\begin{figure}[h]
\includegraphics[angle=0,width=0.8\linewidth]{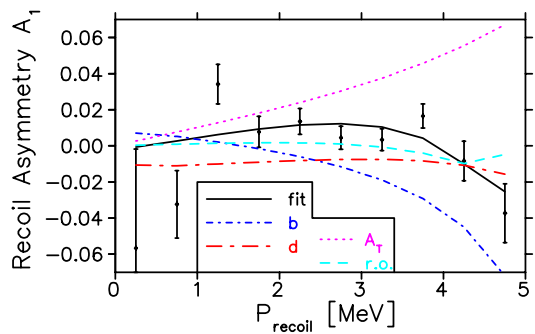}
\caption{ (Color online)
The dependence on recoil momentum of the recoil asymmetries for the best
fit with $C_T+C_T'$=0 (i.e. assuming the result of Ref.~\cite{carnoy}), 
and $b_M$, $d$ and $C_T-C_T'$ floating. The fit has $\sqrt{\chi^2/N}$= 1.41, 
and the result is $|(C_T-C_T')/C_A|$ = 0.42$^{+0.15}_{-0.32}$ at 90\%
confidence. 
}
\label{left0plot}
\end{figure}

The breakdown into the various contributions is shown in Fig~\ref{left0plot}.
Here it can be seen that $b_M$ and $A_T$ produce contributions to $A_1$ with
similar momentum dependence, so again it is difficult to fit them both
simultaneously. It also is clear the the contribution to 
$A_1$ from $d$ is comparatively 
constant with momentum, so $d$ can be floated in
a meaningful fashion.

\subsection{Constraining $C_T+C_T'$=0 and constraining $b_M$ from theory}

Here we keep $C_T+C_T'$=0 from the literature, 
and also constrain the weak magnetism term
$b_M/(AM_{GT})$ from other physics. 
We assume it is given by the nucleon value 4.7, with arbitrary error
given by the full value 4.7. As we discussed in Section~\ref{recoilorder}, 
there are a number of cases in the literature where the other matrix
element contributing to $b_M$ is not very large, and this range of $b_M$ easily
covers all the cases in the literature.

\begin{figure}[h]
\includegraphics[angle=0,width=0.8\linewidth]{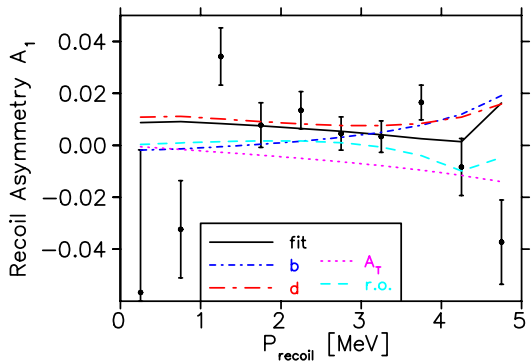}
\caption{ (Color online)
The dependence on recoil momentum of the recoil asymmetries for the best
fit with $C_T+C_T'$=0 (i.e. assuming the result of Ref.~\cite{carnoy}), 
but $b_M/(AM_{GT})$=4.7$\pm$4.7, with $d$ and $C_T-C_T'$ floating.
The result is $A_T$=0.015$\pm$0.029$\pm$0.019 (see text). 
}
\label{left0bfixplot}
\end{figure}

We list in Table~\ref{table:results} the fit results for $A_T$ and $d$ 
for different values of $b_M$, in hopes that further knowledge of $b_M$ and
$d$ will eventually become available.
We show the contributions to the fit from the different terms 
in Fig.~\ref{left0bfixplot}.

\begin{table}[htb]
\begin{tabular*}{3.4in}{crrc}
$b_M/(AM_{GT})$ &     $d/A$~~  &       $A_{T}$~~~~~~  &   $\sqrt{\chi ^2/N}$ \\
\hline \\
9.4  &   -34 $\pm$  42 &   0.034  $\pm$       0.032 &  2.06 \\
4.7  &   -24 $\pm$  39 &   0.015  $\pm$       0.029 &  1.90 \\
0    &   -14 $\pm$   36 &   -0.003 $\pm$       0.027 &   1.76 \\
-4.7 &   -4  $\pm$   34 &  -0.022 $\pm$      0.025 &  1.64   \\ 
-9.4 &   6  $\pm$  32 &   -0.040 $\pm$      0.024 &  1.56   \\ 
-14.1 & 16 $\pm$ 31   &   -0.059 $\pm$ 0.023  & 1.51 \\
-18.8 &  26 $\pm$ 31   &   -0.078 $\pm$ 0.023  & 1.50 \\ 
-23.5 &  37 $\pm$ 31   &   -0.096 $\pm$ 0.024  & 1.52 \\ %
\hline \\
-18$\pm$11 &   24  $\pm$  41 & -0.074 $\pm$ 0.050 &  1.60    \\
\hline 
\end{tabular*}
\caption{Results for fits to the dependence of the recoil asymmetry on recoil
momentum for  $d/A$ and $A_{T}$, with $N=8$. 
 $b_M/AM_{GT}$ is fixed at each of the values in the first column.
We include the nucleon value of $b_M/(AM_{GT})=4.7$, and sweep through a large
number of other values.
The bottom line is the fit result if $b_M$ is allowed to float unconstrained;
note that this value for $b_M$ is far outside the nucleon value.
}
\label{table:results}
\end{table}

We interpret the results of this table to mean 
$A_{T}$= 0.015 $\pm$ 0.029 (statistical) $\pm$ 0.001 (systematic) 
$\pm$ 0.019 (theory syst). The first systematic error is from the nuclear
polarization. 
The second systematic error is dominated by the error on $b_M/(AM_{GT})$ of
100\% of the nucleon value 4.7.

Then, by using Eq.~\ref{eq:Arecoil}, 
we can interpret our result for $A_{\rm recoil}$ to imply that 
$C_T C_T'/C_A^2$ = 
0.012 $\pm$ 0.022 $\pm$ 0.015.
We show the resulting 90\% confidence limits in Fig.~\ref{Rb:exclusionLR} as
the green horizontal-hashed rectangular exclusion region
showing $|(C_T-C_T')/C_A| \leq 0.36$ (with $C_T + C_T'$ assumed 0).
It can be seen that under these assumptions,  we can place constraints on 
tensor interactions coupling to non-Standard Model right-handed neutrinos,
i.e. on $C_T -C_T'$,
competitive with constraints derived from the
$\beta$-$\nu$ correlation in $^{6}$He~\cite{ornl,gluck}.

\subsection{Future improvements}

In order to improve the precision of this experiment,
2nd-order recoil terms that might contribute to the 1$^+$ to 2$^+$ transition 
would have to be addressed.
By adding efficient $\gamma$-ray detection to  
measure the recoils in coincidence with
the 617 keV $\gamma$ ray, the experimental asymmetry for the excited-state
transition could be measured separately. Then it could be included 
in the model for the total asymmetry. 
That would allow the extraction of new physics from the 1$^+$ to 0$^+$ 
transition free of
the complications of the higher-order recoil effects of the excited state
transition. This would be necessary to improve the sensitivity
to below the 0.01 level. Other improvements would include implementing the
better optical pumping used in Ref.~\cite{melconian}, which would improve
accuracy by almost a factor of 2 by making the polarization close to 1.0. 

An accurate simultaneous 
$\beta$ asymmetry experiment as a function of $\beta$ momentum
could also help constrain the recoil-order terms.

Possible extensions of this experiment would include measuring the
same quantity in $^{82}$Rb~\cite{vieira}
as a nuclear structure consistency test in the same shell.
The ground state transition is an 82\% branch, and the lower Q-value 
(3.4 vs. 4.7 MeV) also would help reduce dependence on recoil-order terms. 
The 3$^+$ to 2$^+$ decay of the $^{38}$K ground state could also be used
to search for tensor interactions---
although several 2nd-order recoil nuclear 
matrix elements are needed, calculations in the SD shell could be done
with reliable estimates of the theoretical error. 
Recoil asymmetry measurements in $^8$Li would require reconstruction of the
momentum of the $\alpha$ particles emitted, and could contribute to 
searches for
2nd-class tensor interactions~\cite{wilkinson}
along with fundamental tensor interactions.  

\section{Conclusions}

We have used atom trap technology to make the first measurements of
the asymmetry of daughter nuclei with respect to the nuclear spin, as suggested
by Treiman~\cite{treiman}. By measuring the momentum dependence of the 
asymmetry, we can  
constrain the recoil order induced tensor $d$ independently. 
The similar momentum dependence of weak magnetism $b_M$ and the 
non-SM tensor physics
$b_T$ and $A_T$ produces correlations in their extraction.
So in Section~\ref{results}
above we have considered the
constraints on non-SM tensor interactions from the present experiment, 
using different assumptions from other
experiments and from theory. 

If we make no assumption about tensor parameters 
from other experiments, and do not constrain $b_M$
and $d$, then the 90\% CL constraints are shown in 
Fig.~\ref{Rb:exclusionLR}.
These are not competitive for tensors coupling to SM left-handed neutrinos, 
but 
their intersection 
with the vertical $(C_T+C_T')=0$ axis 
shows the potential sensitivity of the technique. 

We believe our most useful result is to 
assume that tensors coupling to SM left-handed neutrinos do not 
exist (i.e. $C_T+C_T'=0$) in accordance with Ref.~\cite{carnoy}, and let $d$ 
float while constraining $b_M/(AM_{GT})$ to the nucleon value with 100\% error
at one sigma.
Then we extract $A_T$ = 0.015 $\pm$ 0.029 (stat) $\pm$ 0.019 (syst).
Eq.~\ref{eq:Arecoil} then implies $C_T C_T'/C_A^2$ = 
0.012 $\pm$ 0.022 $\pm$ 0.015.
The resulting constraints on tensor couplings to right-handed neutrinos 
$|(C_T-C_T')/C_A|<0.36$ are also indicated in Fig.~\ref{Rb:exclusionLR}. 
These constraints are complementary to those from the $^{6}$He $\beta$-$\nu$ 
correlation. 
We include Table I in hopes that future improvements in the knowledge of
the recoil-order terms can be included.

So our results place constraints on tensor interactions complementary to those
from other experiments. 
The systematic error is dominated by uncertainty in extracting 
recoil order corrections from our data, and there are no serious
experimental systematics.
The statistical 
error could be made considerably smaller with more counting and modest
experimental improvements, which would cut the overall error by more than
a factor of two even without theoretical guidance, and possibly by the order of
magnitude that would make it complementary to the best measurements in 
nuclear, neutron, and pion decay.

\begin{acknowledgements}
Supported by the Natural Sciences and Engineering Council of Canada, 
National Research Council Canada through TRIUMF, WestGrid, 
and the Israel Science Foundation.
\end{acknowledgements}



\appendix
\section{}

We present here the analytic results for the recoil spin asymmetry as a function of recoil momentum, assuming no Fermi function.
These are obtained from the expressions for the full angular distribution
in Refs.~\cite{jtw} and~\cite{holstein} 
by integration over $\nu$ and $\beta$ momenta. 
See Ref.~\cite{aviv} for more details. The units set $m_\beta$=1.

The angular distribution of daughter nuclei with respect to the 
nuclear spin as a function of their momentum $P_r$ is 

\begin{eqnarray*}
\lefteqn{W[P_r,\theta] dP_r d(\cos{\theta_r})}\\
 & = &  
[f_1(P_r) + b_{T} f_6(P_r) - (a_{\beta \nu} + \frac{cT}{3})f_2(P_r) + cT f_3(P_r) 
\\
&  &
-PA_{T}f_4(P_r){\rm cos}(\theta_r) + cT f_5(P_r){\rm cos}^2(\theta_r) 
\\
& & 
+P b_T f_7(P_r) {\rm cos}(\theta_r)] dP_r d(\cos{\theta_r}). 
\end{eqnarray*}

\noindent 
Integration of these kinematic functions over recoil momentum 
$P_r$ produces the terms in Eq.~1.
We also include here the normalization effect of the Fierz interference term 
$b_{T}$ and a similar effect $b_T$ in the asymmetry term, each
multiplied by $m_\beta/E_\beta$ before integration over $\beta$ energy. 
These terms are explicitly neglected in 
Eq.~1 and in Ref.~\cite{treiman}.

Note that the effect of the normalization
term scaling $f_6$ is neglible in this work, because in Gamow-Teller 
decays $A_{\rm recoil}$=0 in the absence of tensor terms and the effects
of $f_6$ enter in higher order in the small tensor parameters.
That is no longer true for mixed Fermi/Gamow-Teller transitions. 

The kinematic functions of momentum are given by

\begin{eqnarray*}
f_1(P_r) & = & \\ 
\lefteqn{\frac{(P_r-E_0^2P_r+P_r^3)^2(3E_0^4+P_r^2+P_r^4+E_0^2(3-4P_r))}{12(E_0^2-Pr_r^2)^3}~~~~~~~~~~~~}
\\ 
f_2(P_r) & = &  \\ 
\lefteqn{\frac{(P_r-E_0^2P_r+P_r^3)^2(3E_0^4+P_r^2(5P_r^2-1)-E_0^2(3+8P_r^2))}{12(E_0^2-Pr^2)^3}~~~~~~~~~~~~}
\\
f_3(P_r) &  = &  -\frac{P_r^2(1-E_0^2+P_r^2)^3}{12(E_0^2-Pr_r^2)^2} \\
f_4(P_r) & = &
\frac{E_0P_r^3(2+E_0^2-P_r^2)(1-E_0^2+P_r^2)^2}{6(E_0^2-P_r^2)^3} \\
f_5(P_r) & = & -\frac{P_r^4(2+E_0^2-P_r^2)(1-E_0^2+P_r^2)^2}{6(E_0^2-P_r^2)^3}~~~~~~~~~~~~~~~\\
f_6(P_r) & = & \frac{E_0^3P_r^2-P_r^4E_0+2E_0P_r^2}{2(E_0^2-P_r^2)}
\\
f_7(P_r) & = & \frac{(E_0^2+P_r^2+1)(E_0^2-P_r^2-1)-E_0^2 P_r (E_0^2-P_r^2)}
{2 (E_0^2-P_r^2)}
\end{eqnarray*}

\noindent 
where $E_0$ is the maximum total energy of the $\beta$, and 
where energies and momenta are in units of $m_{\beta}$. Integrating over the
daughter nucleus momentum produces the terms needed in Eq.~1:

 \begin{eqnarray*}
 x_1 & = &
 \frac{5(E_0^5-6E_0^3+3E_0+\frac{2}{E_0}+12E_0lnE_0)}{4\sqrt{E_0^2-1}(2E_0^4-9E_0^2-8)+A_{ln}} \\
 x_2 & = &
 \frac{\sqrt{E_0^2-1}(4E_0^4-28E_0^2-81)+15(6+\frac{1}{E_0})A}{4\sqrt{E_0^2-1}(2E_0^4-9E_0^2-8)+A_{ln}} \\
A_{ln} & = & 
\ln{(E_0 + \sqrt{E_0^2 - 1})}
  \end{eqnarray*}

Note the typographical error in the equation for $x_1$ in Ref.~\cite{treiman}.

\end{document}